\documentclass[aps,prd,preprint,superscriptaddress,tightenlines,nofootinbib]{revtex4}

\usepackage{graphicx}
\usepackage{dcolumn}
\usepackage{bm}

\def\etal   {{\it et al.}}

\begin{document}

\preprint{CLNS 04/1900}       
\preprint{CLEO 04-18}         

\title{The Search for $\eta(1440) \to K^0_S K^\pm \pi^\mp$ in Two-Photon Fusion at CLEO}

\author{R.~Ahohe}
\author{D.~M.~Asner}
\author{S.~A.~Dytman}
\author{W.~Love}
\author{S.~Mehrabyan}
\author{J.~A.~Mueller}
\author{V.~Savinov}
\affiliation{University of Pittsburgh, Pittsburgh, Pennsylvania 15260}
\author{Z.~Li}
\author{A.~Lopez}
\author{H.~Mendez}
\author{J.~Ramirez}
\affiliation{University of Puerto Rico, Mayaguez, Puerto Rico 00681}
% ...
\collaboration{CLEO Collaboration} 
\noaffiliation

\date{January 10, 2005}

\begin{abstract} 
We analyze 13.8 ${\rm ~fb^{-1}}$ of the integrated $e^+e^-$ 
luminosity collected at 10.6 GeV center-of-mass energy with 
the CLEO~II~and~CLEO~II.V detectors 
to study exclusive two-photon production of hadrons 
with masses below $1.7{\rm ~GeV/c^2}$ 
decaying into the $K^0_S K^\pm \pi^\mp$ final state. 
We observe two statistically significant enhancements 
in the $\eta(1440)$ mass region. These enhancements 
have large transverse momentum which rules them out  
as being due to pseudoscalar resonances but is consistent 
with the production of axial-vector mesons. 
We use tagged two-photon events to study the properties of the 
observed enhancements and associate them with 
the production of $f_1(1285)$ and $f_1(1420)$. 
Our non-observation of $\eta(1440)$ is inconsistent 
by more than two standard deviations 
with the first observation of this resonance 
in two-photon collisions by the L3 experiment. 
We present our estimates for 90\%~confidence level upper limits on the products of two-photon 
partial widths of pseudoscalar hadrons 
and their branching fractions into $K^0_S(\pi^+\pi^-)K^\pm\pi^\mp$. 
\end{abstract}

%
%
\maketitle


\section{Introduction}
\label{introduction}

A key to understanding the phenomenon of quark and gluon confinement in 
Quantum Chromodynamics (QCD) is the experimental observation and 
analysis of the properties of various hadronic bound states 
predicted by the quark model, Lattice QCD (LQCD),  Flux Tube 
and other theories\cite{morningstar,bernard}. 
One of the main emphases in hadronic physics has long been 
on the discovery of exotic hadronic resonances and ``extranumerals'' 
which could not be explained within the framework of the quark model. 
For example, the Flux Tube model and LQCD (even when the quenching approximation is lifted) 
predict a large number of light glueballs -- bound states of the carriers 
of strong interaction -- and of hybrids -- hadrons composed of three constituents, 
two quarks and a gluon. 
A large number of potential candidates for such new states of hadronic matter
have been discovered over the past 35 years in many experiments\cite{PDG}.
In our opinion, most of these candidates need to be confirmed and remain
to be understood. 
One such candidate is the $\eta(1440)$ resonance 
first observed in 1967 in $p\bar{p}$ annihilation at rest into $K\bar{K}\pi\pi^+\pi^-$.
This resonance has also been seen in radiative decays of $J/\psi$
into $K\bar{K}\pi$ and in charge-exchange hadronic reaction $\pi^-p \to \eta \pi\pi n$.
Until recently, the $\eta(1440)$ had been seen only in gluon-rich environments, 
and this established it as a prominent glueball candidate. 
Alternatively, $\eta(1440)$ and another mysterious pseudoscalar hadron, $\eta(1295)$, 
might be radial excitations of the $\eta$ and $\eta^\prime$ mesons. 

One way to discriminate among 
the ground state meson, radial excitation, and glueball hypotheses 
is to measure a hadron's two-photon partial width, $\Gamma_{\gamma\gamma}$. 
Assuming that quantum numbers allow a two-photon decay, 
$\Gamma_{\gamma\gamma}$ 
would be of the order keV for a ground state meson, 
approximately an order of magnitude smaller for a radial excitation\cite{page}, and 
of a vanishingly small value for a true glueball, because photons couple to gluons 
only through an intermediate quark loop. A two-photon partial width is usually 
measured using inverse two-photon decay, {\it i.e.}, in the process of single 
hadron production in $e^+e^-$ scattering, where the hadron is born in the fusion of 
two space-like photons emitted by the beam electron and the positron. 
It is also possible that light glueballs and pseudoscalar mesons, 
such as the $\eta(1440)$, are mixed, and their parameters, 
including $\Gamma_{\gamma\gamma}$, 
should be obtained from global fits to the light hadron spectrum. 
In 2001 a new piece was added to the $\eta(1440)$ puzzle, when
the L3 experiment reported\cite{L3} the first observation of the $\eta(1440)$
in two-photon collisions. L3 measured $\eta(1440)$'s two-photon partial width
to be $212 \pm 50~{\rm (stat.)}\pm 23~{\rm (sys.)}$~eV, assuming 
it decays only to $K\bar{K}\pi$. The other poorly understood hadron decaying to $K \bar{K} \pi$, 
$\eta(1295)$, has still not been observed in two-photon collisions. 

In this paper we seek corroborating evidence 
for the L3's observation of the $\eta(1440)$. Using a data sample that exceeds 
that of L3 by a factor of five, we do not observe this resonance. 
In our analysis, we study the reaction $e^+e^- \to e^+e^- {\cal R}$, 
where the hadron ${\cal R}$ (of mass below $1.7~{\rm GeV/c^2}$) 
is produced by two space-like photons, 
decays to $K\bar{K}\pi$, 
and 
is reconstructed in its decay to 
the final state $K^0_S K^\pm \pi^\mp$.\footnote{
Charge-conjugated states are implied throughout this paper.} 
The secondary vertex associated with the decay $K^0_S \to \pi^+\pi^-$ 
helps to identify $K \bar{K} \pi$ production. The presence of 
four charged final state particles helps to trigger on such events. 
The quantity that allows us to distinguish between two-photon and other 
production processes for pseudoscalar mesons is the transverse momentum $p_\perp$, 
which is the component of the hadronic system's 
total vector momentum 
perpendicular to the beam axis. 
The properties of the $e^+e^-$ scattering amplitude demand two-photon 
events peak sharply at small $p_\perp$, except when a single axial-vector 
meson, 
such as $f_1(1285)$, $f_1(1420)$, or $f_1(1510)$ 
is produced. 
The production of these mesons in inverse two-photon decay is suppressed 
by the Landau-Yang theorem\cite{landau} for real photons. Therefore, the production 
cross sections for axial-vector mesons are enhanced at intermediate $p_\perp$. 
Finally, most of the time, the scattered electron and positron 
carry a large fraction of the event energy 
away in the direction of the beam 
and are not detected, 
{\it i.e.}, remain ``untagged''. 
When the scattered electron or positron is detected (``tagged'') 
in the calorimeter, the tag and the hadronic system are 
recoiling against each other and, most of the time, 
their momenta are collinear in the plane perpendicular to the beam axis. 

In our analysis, we study the $K^0_S K^\pm \pi^\mp$ final state 
in two regimes: in the untagged mode with $p_\perp$ below 0.6~GeV/c and 
in the tagged mode where $p_\perp$ usually exceeds 1.0~GeV/c. 
Taking into account the integrated $e^+e^-$ luminosities 
and two-photon production cross sections for pseudoscalars, 
our event sample in the $\eta(1440)$ mass region exceeds that of 
the L3 experiment by a factor of five. 
We reported preliminary results of our analysis in the untagged mode previously\cite{panic02}. 

This paper is organized as follows: 
first, we briefly describe our detector, data sample, 
and our two-photon Monte Carlo generators. Then, we present the analysis 
procedure and the calibration data sample. 
The section on our results for the pseudoscalar production 
and systematic errors is followed by the section 
describing the analysis of axial-vector mesons and the conclusions. 
The main result of our study is the non-observation of the $\eta(1440)$ 
in our data sample where we expected $114 \pm 28$ such events 
according to the results published by the L3 experiment\cite{L3}. 

\section{Experimental Apparatus and Data Sample} 
\label{apparatus} 

The results presented here 
are obtained from data 
corresponding to 
an integrated $e^+e^-$ luminosity of
$13.8 {\rm ~fb^{-1}}$. 
The data were collected 
at the energies of the $\Upsilon(4{\rm S})$ resonance mass 
and 60~MeV below it 
at the 
Cornell Electron Storage Ring (CESR) 
with the CLEO detector. 
The first third of the data was recorded with
the CLEO~II configuration of the detector\cite{CLEOII_description},
which consisted of three cylindrical drift chambers
placed in an axial solenoidal magnetic field of 1.5T,
a CsI(Tl)-crystal electromagnetic calorimeter,
a time-of-flight (TOF) plastic scintillator system
and a muon system (proportional counters and 
copper strips embedded 
at various depths in steel absorbers).
Two thirds of the data were taken with
the CLEO~II.V configuration of the detector
where the innermost drift chamber
was replaced by a silicon vertex detector\cite{CLEOII.V_description} (SVX)
and the argon-ethane gas of the main drift chamber
was changed to a helium-propane mixture.
This upgrade led to improved resolutions in momentum
and, to lesser extent, in specific ionization energy loss
($dE/dx$) measurements.

When the data were collected, 
the information from the two outer drift chambers,
the TOF system, and electromagnetic calorimeter 
was used to make the decisions in the 
three-tier CLEO trigger system\cite{CLEOII_trigger}  
complemented by a software filter for beam-gas event rejection. 
The response of the detector is modeled with 
a Monte Carlo (MC) program 
based on the GEANT~3 simulation framework\cite{GEANT}. 
The data and simulated samples are processed 
by the same event reconstruction program. 
Whenever possible, the efficiencies are either 
calibrated or corrected for the difference 
between simulated and actual detector responses 
using direct measurements from independent data samples. 

\section{Monte Carlo Generators for Two-Photon Events}
\label{mc_generators}

\begin{figure}
\includegraphics[width=6.5in]{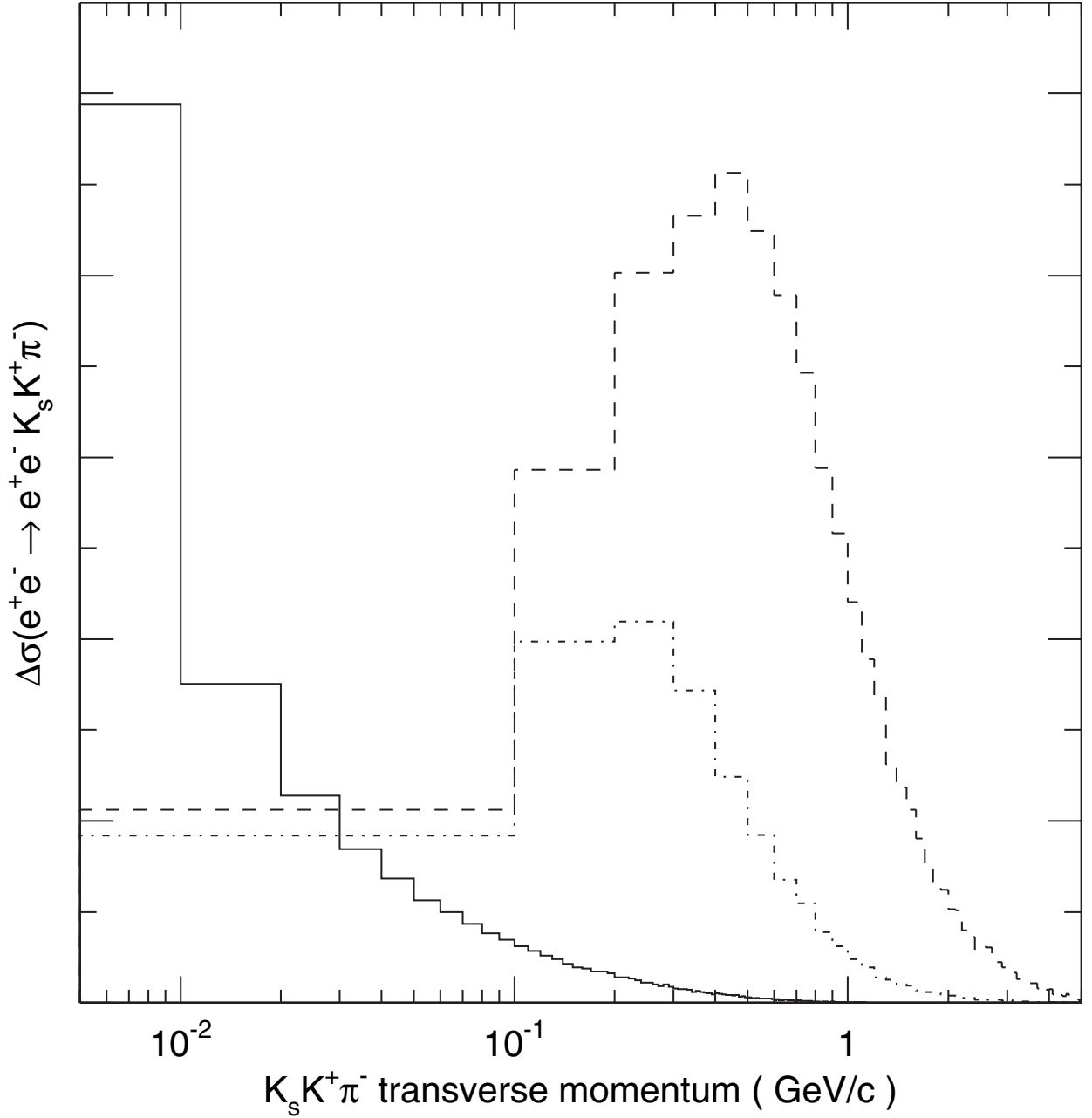}
\caption
{
The shapes of partial 
({\it i.e.} integrated over $p_\perp$ bins) 
cross sections 
$\Delta \sigma (e^+e^- \to e^+e^- {\cal R})$ 
for 
pseudoscalar (solid line, pole-mass parameter 770~MeV) 
and 
axial-vector (dashed and dashed-dotted lines 
for pole-mass parameters of 770~MeV and 400~MeV, 
respectively) mesons in MC at generator level. 
The distributions for axial-vector mesons 
are normalized to the same integrated luminosity. 
Note that a relatively small fraction 
of pseudoscalar events is expected 
in the region of $P_\perp$ above 100~MeV/c, 
where most axial-vector mesons are produced. 
}
\label{fig_mc_ptran_1}
\end{figure}

To lowest order in perturbation theory, the amplitude 
for hadron production in the process of two-photon fusion 
can be factorized into two terms. 
The first term describes the emission of space-like photons 
by the electron and the positron and is completely calculable 
in QED. The second term deals with the production 
of a single $C$-even hadron by these photons. 
This, essentially {\it hadronic} part of the amplitude depends on 
the structure of a particular hadronic resonance and 
can be parametrized in terms of its two-photon partial width 
and transition form factors. 
To describe the factorization of the amplitude 
and 
to relate 
a production cross section 
to 
the two-photon partial width of a hadron, 
we employ a theoretical framework developed independently 
by several theorists and conveniently summarized 
by V.M.~Budnev \etal \cite{ggmc}. 
In this framework, the QED part of the amplitude is calculated 
using helicity tensors for fluxes of incoming photons 
of various polarizations. The hadronic part of the 
amplitude is described by transition form factors 
which are the functions of $Q^2$, 
{\it i.e.} the masses of the space-like photons. 
In our previous analysis\cite{paper_ff}, 
we demonstrated that light pseudoscalar mesons 
have similar transition form factors that 
can be parametrized in terms of pole-mass parameters. 
Helicity and parity conservation laws and Bose symmetry 
imply that only transverse photons interact to produce 
pseudoscalar final states. 
This holds even for highly space-like photons and explains 
why there is only one transition form factor that affects 
the cross section for $e^+e^- \to e^+e^- \eta(1440)$. 
We choose the pole-mass parameter that determines $Q^2$ evolution of 
this transition form factor to be a commonly used value of 770~MeV 
(which is {\it coincidentally} 
close to the $\rho$ meson mass). However, our results for pseudoscalar 
mesons are insensitive to the value of this parameter. 
This is the case because most of the cross section 
for pseudoscalar production comes from the low $Q^2$ region, 
where transition form factors are unimportant.
We show the generator-level distribution of $p_\perp$ 
for the $\eta(1440)$ signal MC in Fig.~\ref{fig_mc_ptran_1}. 
There is a strong correlation between $Q^2$ and $p_\perp^2$ 
in case of pseudoscalar production but only at very small $p_\perp^2$.  
The momentum transfer $Q^2$ cannot be reconstructed unambiguously 
for untagged events, so we use $p_\perp$ 
(which is measured using the hadronic system alone) 
in our analysis. Note that for the axial-vector mesons, 
$Q^2$ can be reconstructed with excellent accuracy 
from the measurements for the tagging electron 
without using the hadronic system. 

Two Monte Carlo generators were developed at CLEO 
for two-photon studies. Both generators are based 
on the same theoretical framework\cite{ggmc} and extensively 
employ an importance sampling technique to simulate the 
divergent QED part of the two-photon amplitudes. 
Our two Monte Carlo generators for pseudoscalar production 
differ mainly by their user interfaces. 
Numerical predictions made using these programs are practically identical, {\it e.g.}, 
both generators predict the $\eta_c$ production cross section at 10.58 GeV 
to be 2.4~pb per 1.0~keV two-photon partial width. 
The generator we use for the analysis described in this paper 
parametrizes wide resonances by relativistic Breit-Wigners 
for two-body decays. 
For narrow resonances, the difference between 
the cross sections evaluated using 
narrow-width approximation 
and 
this more realistic treatment of the running mass 
is approximately 1\%. 
Both generators predict angular distributions 
of the decay products correctly 
and were employed in all previously published 
CLEO publications on two-photon production\cite{cleo_gg_all}. 
We emphasize our high confidence in the quality of our MC generators, 
because their numerical predictions are very important for measuring two-photon 
partial widths and transition form factors. 

Assuming that transition form factors are known, 
the two-photon partial width $\Gamma_{\gamma\gamma}({\cal R})$ 
of the resonance ${\cal R}$ can be measured 
(up to the uncertainty in branching fractions) 
from data 
\begin{eqnarray}
\Gamma_{\gamma\gamma}({\cal R}) ~ {\cal B}(K^0_S (\pi^+\pi^-) K^\pm \pi^\mp ) ~({\rm keV})~= 
\frac{N^{\rm data}}{{\cal L} ~ \epsilon ~ \sigma^{\rm MC} }
\label{equation_1}
\end{eqnarray}
\noindent where $N^{\rm data}$ is the estimate of the number of detected signal events, 
${\cal L}$ is the integrated $e^+e^-$ luminosity of the data sample, 
$\epsilon$ is the overall trigger and reconstruction efficiency (excluding 
branching fractions), and $\sigma^{\rm MC}$ is the numerically predicted 
cross section for the process $e^+e^- \to e^+e^- {\cal R}$ 
with $\Gamma_{\gamma\gamma}({\cal R}) = 1$~keV. 
We evaluate the cross sections $\sigma^{\rm MC}$ numerically using our 
MC generators.  These quantities depend 
on the assumptions about the transition form factors, 
the electron beam energy, the mass and full width 
of the resonance ${\cal R}$, and on the beam polarization, 
although the CESR beams are not polarized. 

Axial-vector mesons have zero two-photon partial width, and 
some additional model input is necessary to approximate 
their transition form factors. 
For our analysis, we implement the model of Cahn\cite{model_cahn} 
where axial-vector mesons are treated 
as non-relativistic $q\bar{q}$ states. 
In this model, the production cross section 
is determined by the normalization parameter $\tilde{\Gamma}_{\gamma\gamma}$ 
and by the transition form factors driven by a single pole-mass parameter.  
In this model, even if $\tilde{\Gamma}_{\gamma\gamma}$ is fixed, 
the pole-mass parameter has a strong impact on the production rate. 
This explains the dramatic difference in the magnitudes of 
cross sections for axial-vector mesons  
shown in Fig.~\ref{fig_mc_ptran_1} 
for pole-mass parameters of 
400~MeV and 770~MeV. 
This feature of Cahn's model for axial-vector 
production cross sections was previously noticed by the TPC/2$\gamma$ 
experiment\cite{tpc2g_f1} when they estimated the strength 
of the $f_1(1420)$ coupling to space-like photons, 
${\cal B}_{K\bar{K}\pi} \tilde{\Gamma}_{\gamma\gamma}=1.3 \pm 0.5 \pm 0.3$~keV, 
by assuming a pole-mass of 770~MeV. The parameter $\tilde{\Gamma}_{\gamma\gamma}$ 
provides an overall normalization in Cahn's model, 
while the differential cross section approaches 
zero, as it should, in the limit $p_\perp^2 \to 0$. 

\section{Analysis Procedure}
\label{analysis_procedure}

In our untagged analysis, we select events with exactly four charged tracks 
that are reconstructed in the region of the detector 
where trigger and detection efficiencies 
are well understood and associated systematic errors are under control. 
We require at least one charged track with 
transverse momentum exceeding 250~MeV/c. The projection of 
this track is required to point to the barrel part of the calorimeter 
({\it i.e.} with $|\cos{\theta}| \le 0.71$, where 
the polar angle, $\theta$, is measured 
with respect to the beam direction). 
We only use events recorded with trigger configurations 
designed to be efficient for events with at least two charged hadrons in the final state. 
The latter three requirements select events recorded 
with a well-understood trigger. 
This is an important step in our analysis, because most 
untagged two-photon pseudoscalar events cannot be triggered on. 
The overall efficiency of these criteria is 9\%. 
The efficiency of our trigger 
(including a software filter developed to reject cosmic, beam-gas and beam-wall events) 
for the $\eta(1440)$ signal is only 43\%. 
Note that, in a fashion typical for two-photon studies, 
we quote the trigger efficiency for events that satisfy 
kinematic selection. This simplifies the evaluation 
of uncertainties in the trigger and reconstruction efficiencies. 
While the trigger efficiency is small, our confidence in knowing 
its relative uncertainty (14\%) is high. This 
confidence is based on dedicated trigger efficiency measurements 
performed for this purpose at CLEO\cite{paper_kk,paper_pp,paper_ff}. 

Furthermore, we need to suppress backgrounds 
arising from non-signal two-photon events, 
$\tau$ pairs and 
one-photon $e^+e^-$ annihilation to hadrons, 
the first two being the most important contributions. 
To suppress such background events, it is useful 
to identify the presence of hadrons with $s$ quarks, 
since many of the background events do not contain them. 
Remaining events often come from 
$\tau$ pairs or two-photon events 
with extra particles (charged or neutral) besides 
the four reconstructed charged hadrons associated with the possible signal. 
Background events where a substantial amount of energy is carried away by the 
undetected particles usually have large missing transverse momentum 
({\it e.g.} in case of 
neutrinos). In background events we often detect photons 
in the calorimeter (especially 
in case of $\tau$ pairs). Non-signal two-photon events 
with some low-momentum particles escaping detection 
could also pose a problem, unless we require $p_\perp$ to be small.  

\begin{figure}
\includegraphics[width=6.5in]{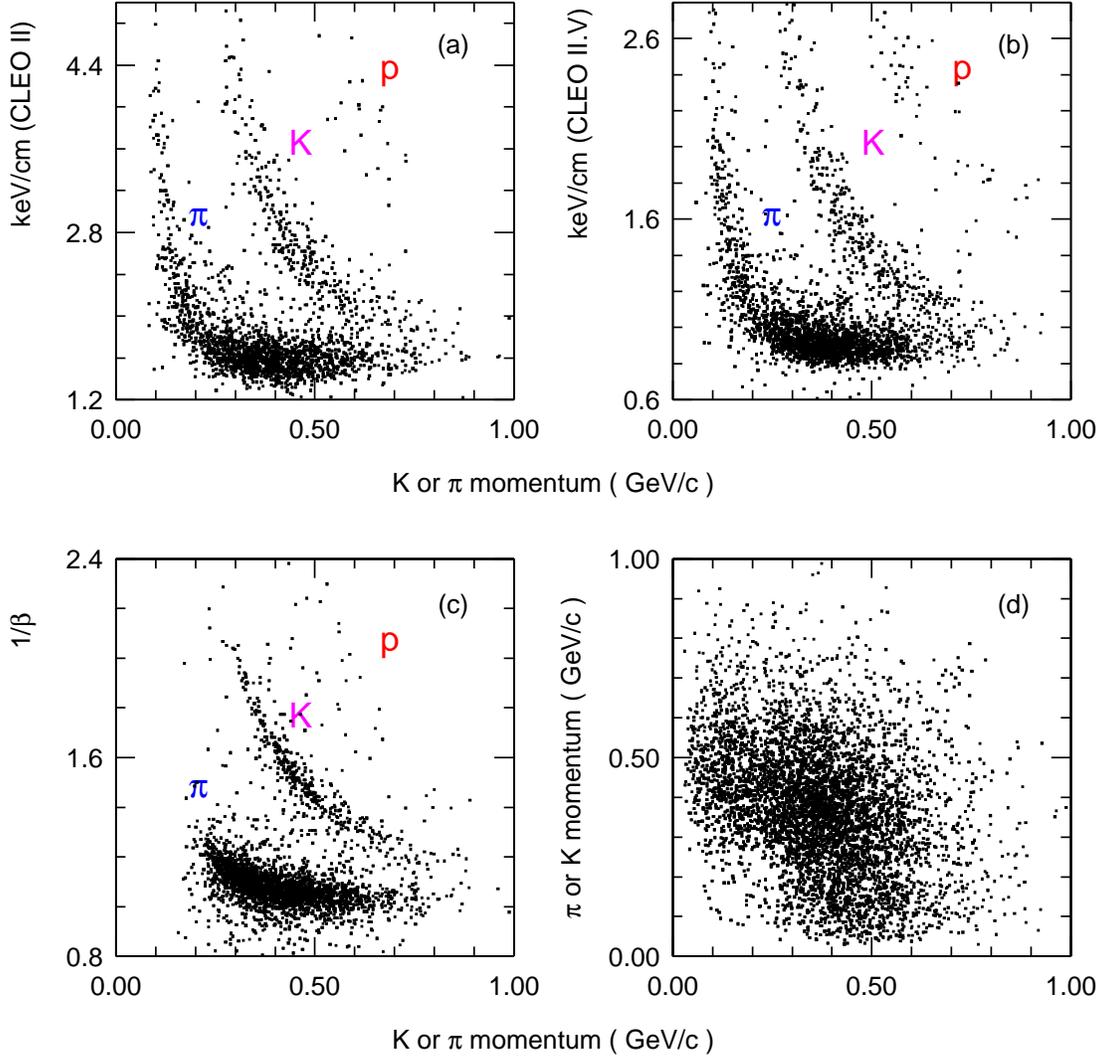}
\caption
{
The measurements of 
$dE/dx$ for (a) CLEO~II and (b) CLEO~II.V, 
and (c) TOF versus charged track momentum 
(shown in (d) before applying the PID requirements) 
for untagged $K^0_S K^\pm \pi^\mp$ data event candidates 
with invariant masses below $1.7 \rm{~GeV/c^2}$ 
after all selection requirements but PID are applied. 
}
\label{fig_data_pid_12}
\end{figure}

To suppress these backgrounds, we identify 
$K^0_S$ candidates by reconstructing 
secondary vertices radially displaced by at least two 
standard deviations ($\sigma$) from the primary interaction point. 
The secondary vertex is required to satisfy 
a set of criteria developed to reduce systematic uncertainty 
in the $K^0_S$ reconstruction efficiency. 
The reconstructed mass of the $K^0_S$ candidate is required to be 
within $\pm5\sigma$ from its nominal value. 
The $K^0_S K^\pm \pi^\mp$ signal-event candidates are required to have only one such $K^0_S$. 
The remaining two charged tracks are identified using 
$dE/dx$ and TOF information. 
The measurements of $dE/dx$ and TOF for these tracks in  
signal-event candidates with $K^0_S K^\pm \pi^\mp$ 
masses below $1.7 {\rm ~GeV/c^2}$ are shown 
in Fig.~\ref{fig_data_pid_12}. 
These measurements are used to 
form a reduced-$\chi^2$ under the different particle identification (PID) 
hypotheses, normalized by the number of available measurements. 
We require the square root of the reduced-$\chi^2$ to be less than three. 
There is negligible PID ambiguity associated 
with this selection for the $\eta(1440)$ candidates in CLEO. 
PID requirements are not imposed on 
the daughter pions from $K^0_S$ candidates in the $K^0_S K^\pm \pi^\mp$ analysis. 
The efficiency of $K^0_S$ and PID selection is approximately 50\%. 
The unmatched neutral energy, 
defined to be the total amount of energy 
collected in photon-like calorimeter clusters 
that do not match the projections of charged tracks, 
is required to be below 100~MeV 
(though one cluster of 1~GeV or more is allowed 
for tagged events, as discussed in more detail later). 
This requirement is powerful in rejecting background from $\tau$ pairs. 
Calorimeter clusters with small amounts of energy could often be present 
in signal events because of ``split-off'' effects caused by 
nuclear interactions of charged hadrons with the materials of the detector. 
The unmatched neutral energy requirement is 65\% efficient. 
Some of the quoted efficiencies depend on the assumptions about the mass, 
full width and the decay dynamics of the $\eta(1440)$. 
We study these dependencies in our analysis. This is discussed later. 

\begin{figure}
\includegraphics[width=6.5in]{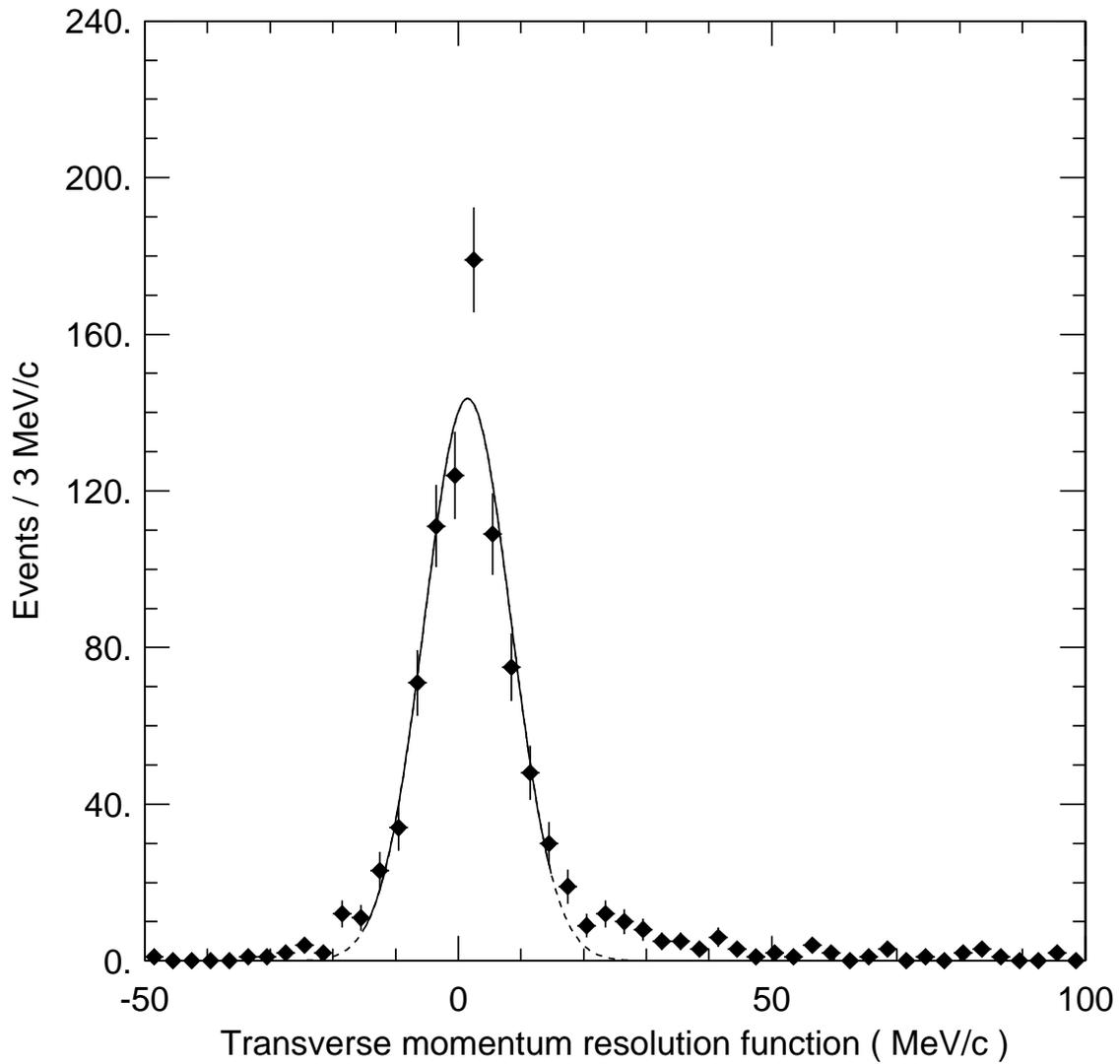}
\caption
{
The difference between the reconstructed and 
the generated scalar values of the $p_\perp$, 
for $\eta(1440) \to K^0_S K^\pm \pi^\mp$ 
signal MC after all selection criteria but $p_\perp$ are applied. 
The curve shows the result of a Gaussian fit to the core part of 
the distribution. 
}
\label{fig_mc_ptran_resolution_13}
\end{figure}

\begin{figure}
\includegraphics[width=6.5in]{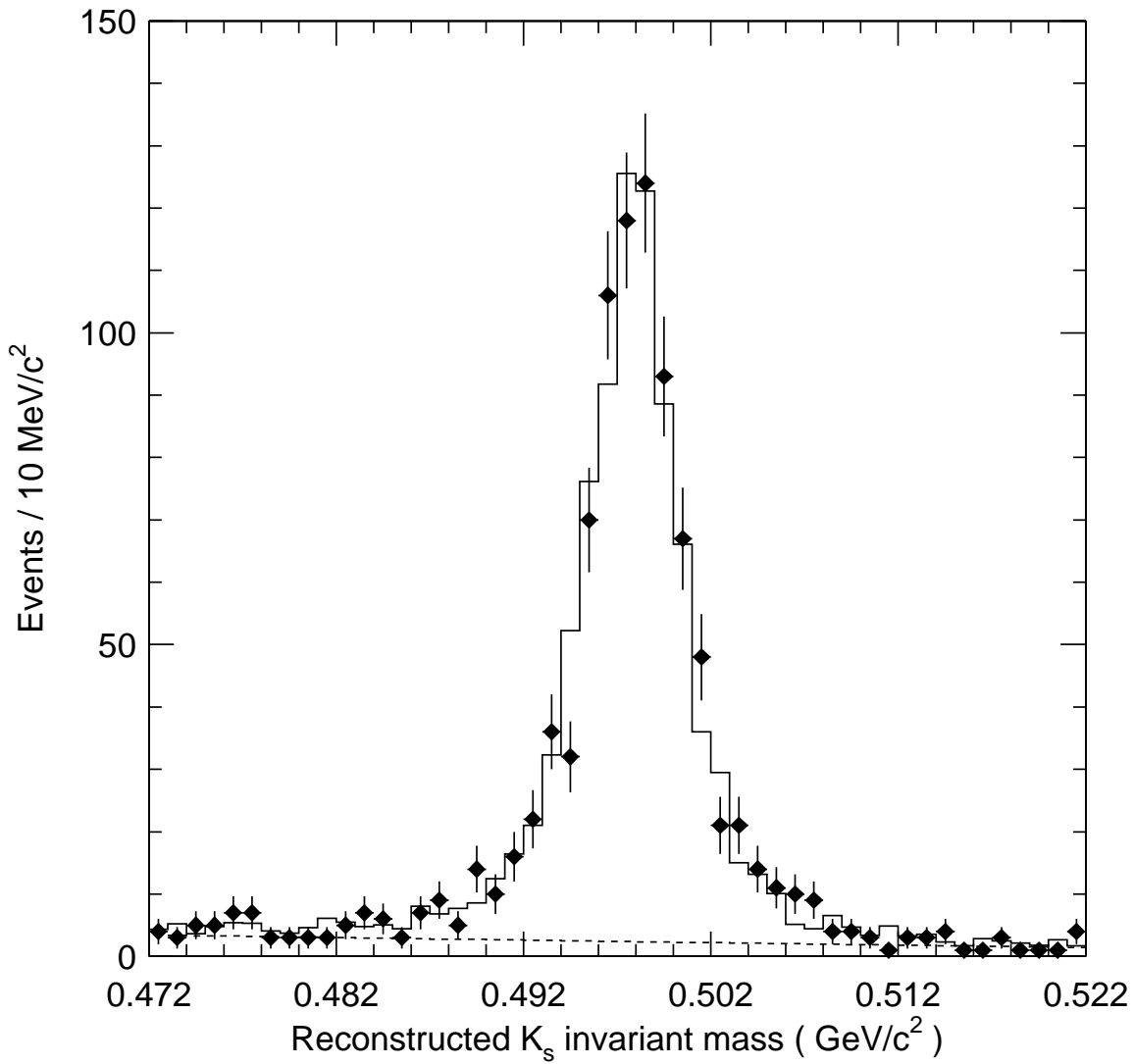}
\caption
{
Reconstructed invariant mass distribution for $K^0_S$ candidates from untagged 
$K^0_S K^\pm \pi^\mp$ data event candidates with mass below $1.7 {\rm ~GeV/c^2}$ 
after all selection criteria are applied. Also shown is the result of 
the fit with $K^0_S$ MC shape and straight-line approximation to the combinatorial background. 
}
\label{fig_data_mass_ks_4}
\end{figure}

In our analysis, $p_\perp$ is the most important quantity 
that discriminates between pseudoscalars born from untagged 
two-photon fusion and other production mechanisms. 
The core part of the $p_\perp$ resolution function is well described by 
a Gaussian shape with $\sigma = 7$~MeV. However, approximately 15\% 
of signal MC events show a $p_\perp$ bias towards larger reconstructed 
values, and the tail stretches to 100~MeV/c. 
We show the difference between the reconstructed and 
generated scalar values of $p_\perp$, 
the $p_\perp$ resolution function, 
in Fig.~\ref{fig_mc_ptran_resolution_13}
for the $\eta(1440)$ signal MC after 
applying all but the $p_\perp$ selection criteria. 
The curve shows the result of a Gaussian fit to the core 
part of the distribution which contains 85\% of the area. 
To suppress combinatorial background, 
untagged events are required to have $p_\perp$ below 100~MeV/c. 
The efficiency of this selection is 65\%. 
Many of the events rejected by this selection are of two-photon origin, 
where some of the final state particles are not reconstructed. 
An example of such feed-down would be the two-photon production 
of final states $K^*\bar{K}^* n(\pi)$ 
(where $n \ge 0$ is the number of pions). 
Cross sections for these processes have never been measured. 
The $p_\perp \le 100$~MeV/c requirement 
strongly suppresses such backgrounds without 
compromising the efficiency for signal events. 
We show the distribution of the $K^0_S K^\pm \pi^\mp$ invariant mass for untagged events 
in Fig.~\ref{fig_data_mass_kskp_7}(a) after all selection criteria are applied. 
The invariant mass distribution for $K^0_S$ candidates (after kinematically constraining 
$K^0_S$ daughters to come from a common secondary vertex) 
from these events is shown in Fig.~\ref{fig_data_mass_ks_4}. 
We discuss our interpretation of Fig.~\ref{fig_data_mass_kskp_7}(a) 
in Section~\ref{results_pseudoscalars}. 

In the tagged event sample, 
in addition to identifying a $K^0_S K^\pm \pi^\mp$ candidate, 
we require the presence of a calorimeter 
cluster with energy of at least 1~GeV and not matching 
the projections of four signal charged tracks into the calorimeter. 
We assume that this cluster is produced by 
the tagging electron or positron which radiated 
the highly off-shell photon in the two-photon process.  
If we find a high-momentum charged track matched to 
this energy cluster, we assume that it is produced by the tag. 
Two-photon tagging is possible on CLEO 
for polar angles larger than $12^\circ$. 
However, at such small angles, the tags barely scrape 
the endcap calorimeter, and tagging efficiency is small. 
For scattering angles between $15^\circ$ and $20^\circ$,  
at least 25\% of the tag energy is usually detected 
(typically, the tagging electron carries at least 4~GeV/c momentum). 
For larger scattering angles, the showers are almost 
fully contained, and the entire energy of the tag is detectable. 
For scattering angles above $22.5^\circ$, calorimeter information 
is complemented by the track reconstruction and independent measurement 
of the tag's momentum. 
For angles above $25^\circ$, the trigger efficiency approaches 100\%. 
A more detailed description of tag identification 
can be found in our previous publication\cite{paper_ff}. 
No transverse momentum requirement is imposed on tagged two-photon events. 

\begin{figure}
\includegraphics[width=6.5in]{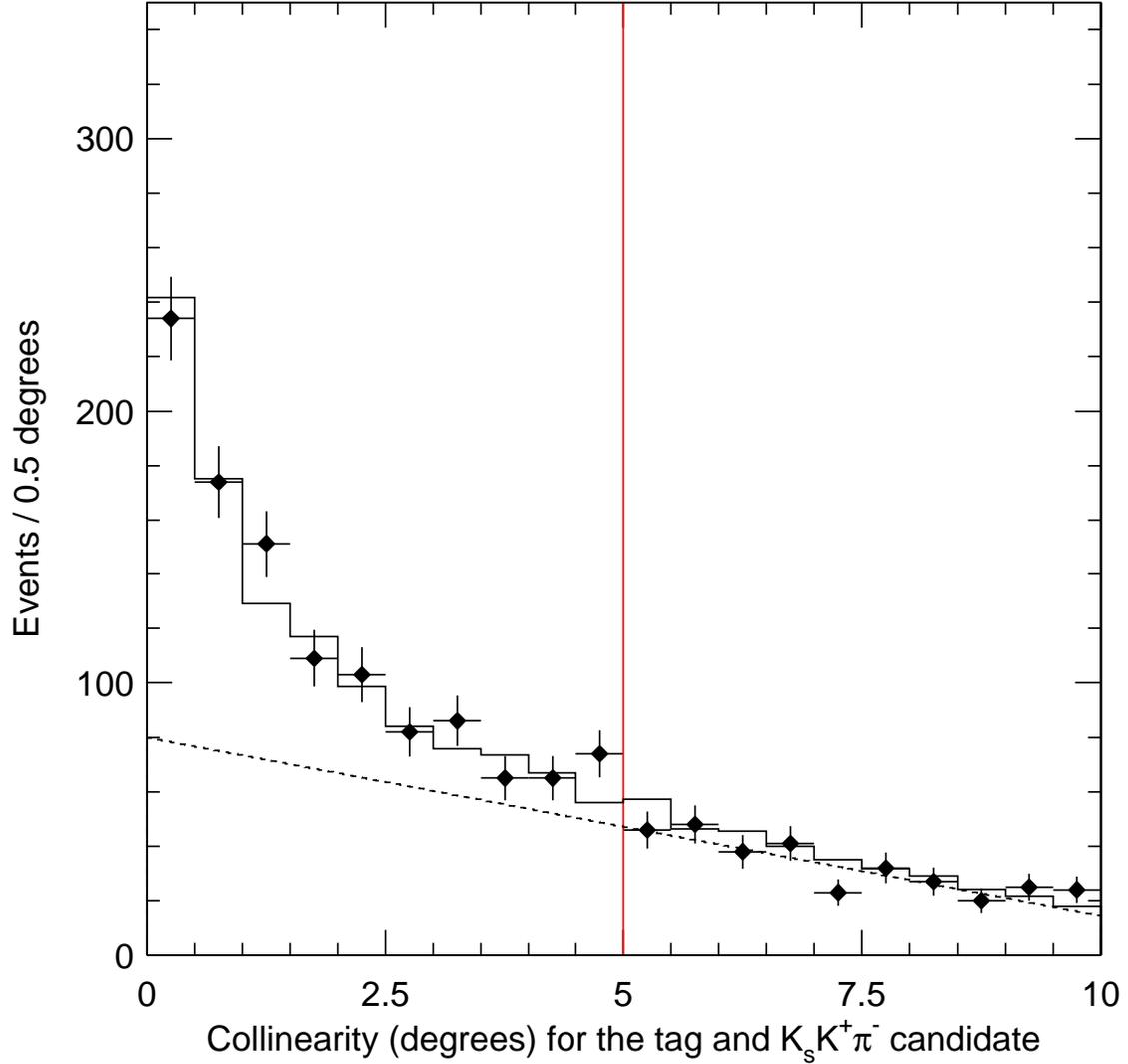}
\caption
{
The collinearity angle (degrees) between 
the $K^0_S K^\pm \pi^\mp$ system and the tagging electron 
(or positron) in the plane perpendicular to the collision axis. 
The data are represented by the points with error bars. 
The dashed straight line approximates 
not-fully reconstructed two-photon events.  
The solid histogram is a fit to the data 
with the signal MC shape for axial-vector meson production 
and the aforementioned approximation for the background.  
The vertical line indicates the selection criterion 
for tagged events. 
}
\label{fig_data_tagged_acol_5}
\end{figure}

To suppress misreconstructed hadronic background, 
we require the tag and the $K^0_S K^\pm \pi^\mp$ candidate 
to be collinear within five degrees in the $(r,\phi)$ plane 
perpendicular to the beam axis. 
The distribution of the collinearity angle, 
which is the deviation of the tag's direction 
from being opposite to that of hadronic system, 
for signal event candidates in data 
is shown in Fig.~\ref{fig_data_tagged_acol_5}. 
This figure also shows the fit to the data with a signal line shape 
from MC simulation for axial-vector two-photon production 
and a linear approximation to the background contribution. 
The shape of the collinearity distribution is determined 
by the QED part of the amplitude and by resolution: while most of the time the 
undetected electron (or positron) transfers some energy to the 
hadronic system, it continues to travel in the direction of the beam. 
This explains the signal shape for the collinearity shown 
in Fig.~\ref{fig_data_tagged_acol_5}. 
The distribution of $K^0_S K^\pm \pi^\mp$ invariant mass 
for tagged events is shown in Fig.~\ref{fig_data_mass_kskp_7}(d). 
We discuss our interpretation of the invariant mass enhancements 
observed in this figure in Section~\ref{results_axial-vectors}. 

\section{Calibration Data Sample}
\label{calibration_data_sample}

The particular values of the selection criteria used in our analysis 
are optimized to provide the best discrimination between 
the signal and background and/or to reduce systematic uncertainties 
in the final results. 
In these optimizations, the expected number of the $\eta(1440)$ signal events 
is estimated from the L3 measurement, 
and the background level is estimated from data 
without inspecting the $\eta(1440)$ signal mass region. 

\begin{figure}
\includegraphics[width=6.5in]{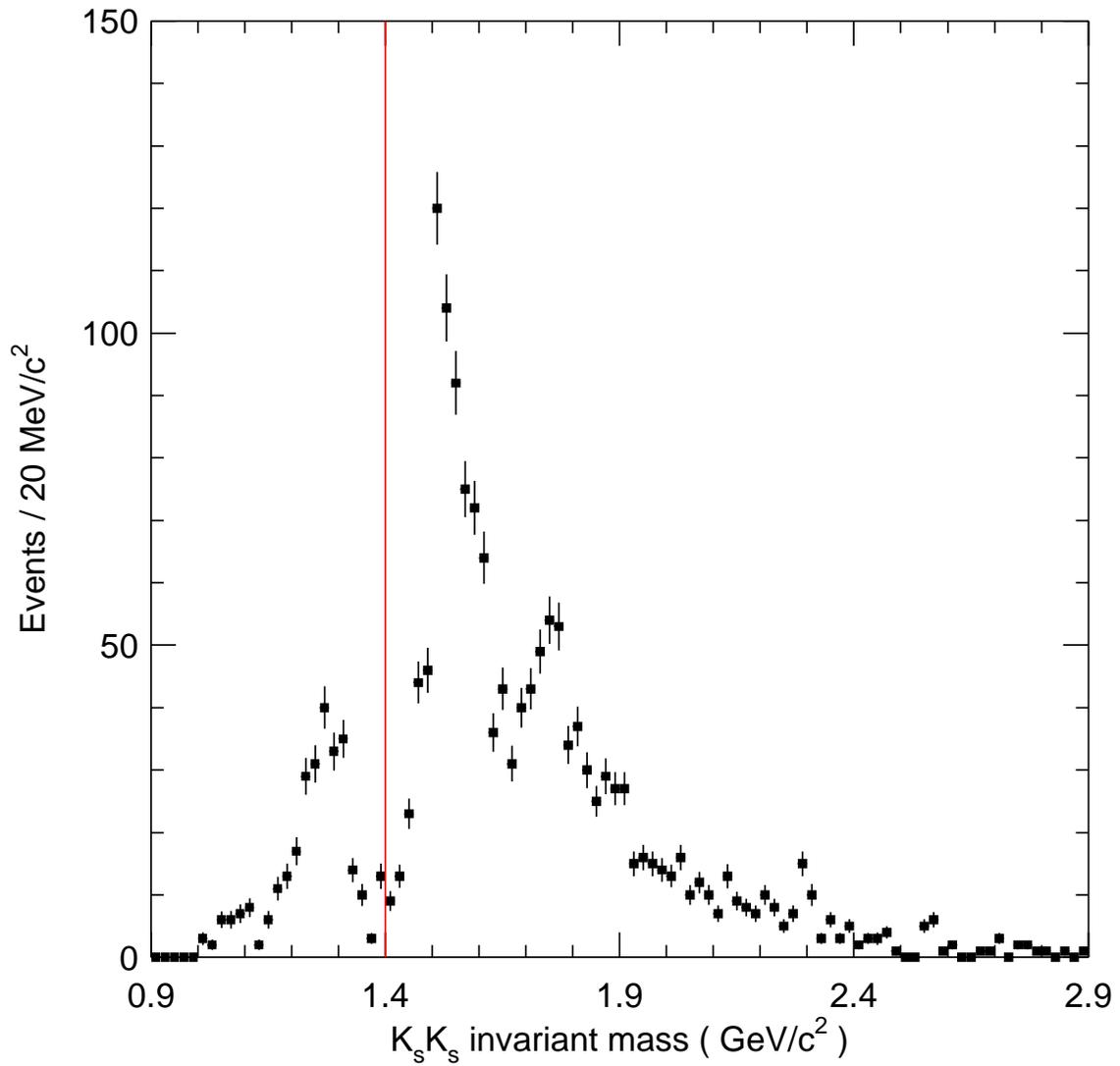}
\caption
{
The invariant mass of $K^0_S K^0_S$ candidates in data. 
Events to the left of the vertical line are accepted 
for calibration and efficiency studies. 
}
\label{fig_data_mass_ksks_2}
\end{figure}

\begin{figure}
\includegraphics[width=6.5in]{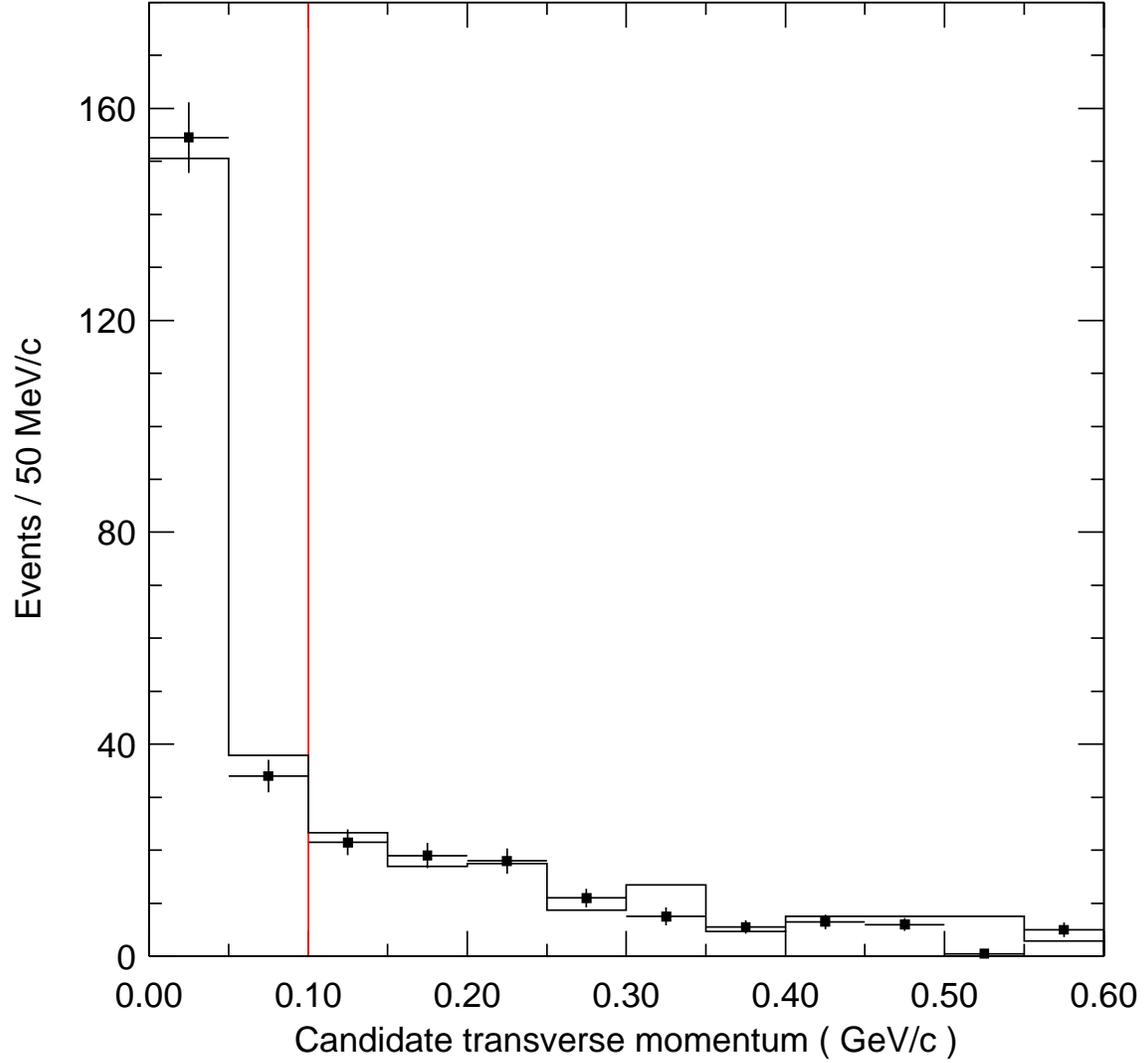}
\caption
{
Transverse momentum of $K^0_S K^0_S$ candidates from the 
calibration data sample (points) 
and MC (solid histogram) after all selection criteria but $p_\perp$ are applied. 
The MC distribution is normalized to data for $p_\perp$ below 100~MeV/c, 
which is also the selection criterion 
applied to our $K^0_S K^\pm \pi^\mp$ candidates 
in the untagged analysis. 
}
\label{fig_data_ptran_ksks_3}
\end{figure}

To measure the efficiencies of our selection criteria 
and to evaluate systematic errors, we use a calibration data sample 
where we find exactly two high-quality $K^0_S$ candidates. 
To establish the two-photon origin for these events, 
daughter pions from the $K^0_S$ pairs are required 
to have a PID-based reduced-$\chi^2 \le 9$. 
The reconstructed masses of the $K^0_S$ candidates 
are required to be 
between $491{\rm ~GeV/c^2}$ and $505{\rm ~GeV/c^2}$. 
We use the distributions of these events' transverse momentum 
and unmatched neutral energy to analyze their 
consistency with the two-photon production mechanism. 
After applying the described selection criteria, 
the $K^0_S K^0_S$ purity of the calibration data sample 
is found to be at least 98\%, {\it i.e.} 
most selected events are due to exclusive 
two-photon production of $K^0_S$ pairs.    
We show the $K^0_S K^0_S$ invariant mass 
for data events selected in this procedure 
in Fig.~\ref{fig_data_mass_ksks_2}. 
To study the efficiencies in the range of 
momenta for final state hadrons from $K^0_S K^\pm \pi^\mp$ signal, 
only $K^0_S K^0_S$ events with invariant mass below 1.4~${\rm GeV/c^2}$ 
({\it i.e.} events to the left of the vertical line in 
Fig.~\ref{fig_data_mass_ksks_2}) 
are used to evaluate the efficiencies and to estimate systematic 
uncertainties for the $K^0_S K^\pm \pi^\mp$ analysis. 
The transverse momentum distributions for our calibration $K^0_S$ pairs 
in data and MC are shown in Fig.~\ref{fig_data_ptran_ksks_3}. 
In this figure, 
to demonstrate that non two-photon contribution is small, 
the $K^0_S K^0_S$ MC sample is normalized to 
the data at transverse momentum values below 100~MeV/c. 
The comparison between MC and calibration data 
indicates no background contribution at 
a statistically noticeable level. 
To study the efficiencies of our selection 
criteria, we loosen and apply them 
individually to the $K^0_S K^0_S$ data and MC samples. 
In each study, we estimate 
the $K^0_S K^0_S$ purity of the calibration data sample 
using the shape of $p_\perp$ for $K^0_S$ pairs. 
We do not find any statistically significant deviations 
between MC efficiencies and their estimates obtained from 
our calibration data sample. The statistics of the latter 
sample are used to estimate systematic uncertainties 
of our selection criteria. 

\section{Upper Limits for Pseudoscalar Production}
\label{results_pseudoscalars}

In our analysis, we study the final state 
$K^0_S K^\pm \pi^\mp$ in two distinct 
kinematic regions: untagged and tagged. 
We use untagged events with $p_\perp \le 100$~MeV/c to 
obtain the results for pseudoscalar mesons. 
Tagged events are used 
to shed light on the untagged high-$p_\perp$ sample. 
We observe no production of the $\eta(1440)$ or any 
other resonance in Fig.~\ref{fig_data_mass_kskp_7}(a) and conclude that 
the two-photon production of $\eta(1440)$ is below our sensitivity. 
The points with the error bars in this figure show our data,  
and the solid curve is the result of a binned 
maximum likelihood (ML) fit described below. 
The dashed curves show the expected 
$\eta(1440)$ signal and its $\pm 1~\sigma$ (statistical) 
deviations according to L3, 
superimposed on the results of our fit. 

To estimate the upper limit on the number of signal events, 
we assume that there is only one resonance potentially 
decaying into $K^0_S K^\pm \pi^\mp$ in the mass region between 
1.3~${\rm GeV/c^2}$ and 1.7~${\rm GeV/c^2}$. 
We perform a binned ML fit to the distribution 
shown in Fig.~\ref{fig_data_mass_kskp_7}(a) with the signal line shape 
for the $\eta(1440)$ and a third order polynomial approximating 
the background contribution. There are several steps involved in 
estimating the signal line shape from MC: 
first, we convolute a two-body relativistic 
Breit-Wigner for a pseudoscalar meson 
with the two-photon luminosity function. 
Then, we correct the resulting function 
with the detector resolution and efficiency functions. 
Using the mass and full width of the $\eta(1440)$ 
obtained by the L3 experiment, 
$M = 1481 \pm 12 {\rm ~MeV/c^2}$ and $\Gamma = 48 \pm 9 {\rm ~MeV}$, 
we expect $114 \pm 28$ signal events in our data. 
From the results of our ML fit, we estimate 
the upper limit on the number of $\eta(1440)$ signal events to be less than 48 
at 90\%~Confidence Level (CL), which is $2.4\sigma$ (stat.) below the prediction 
based on the L3 observation. 

\begin{figure}
\includegraphics[width=6.5in]{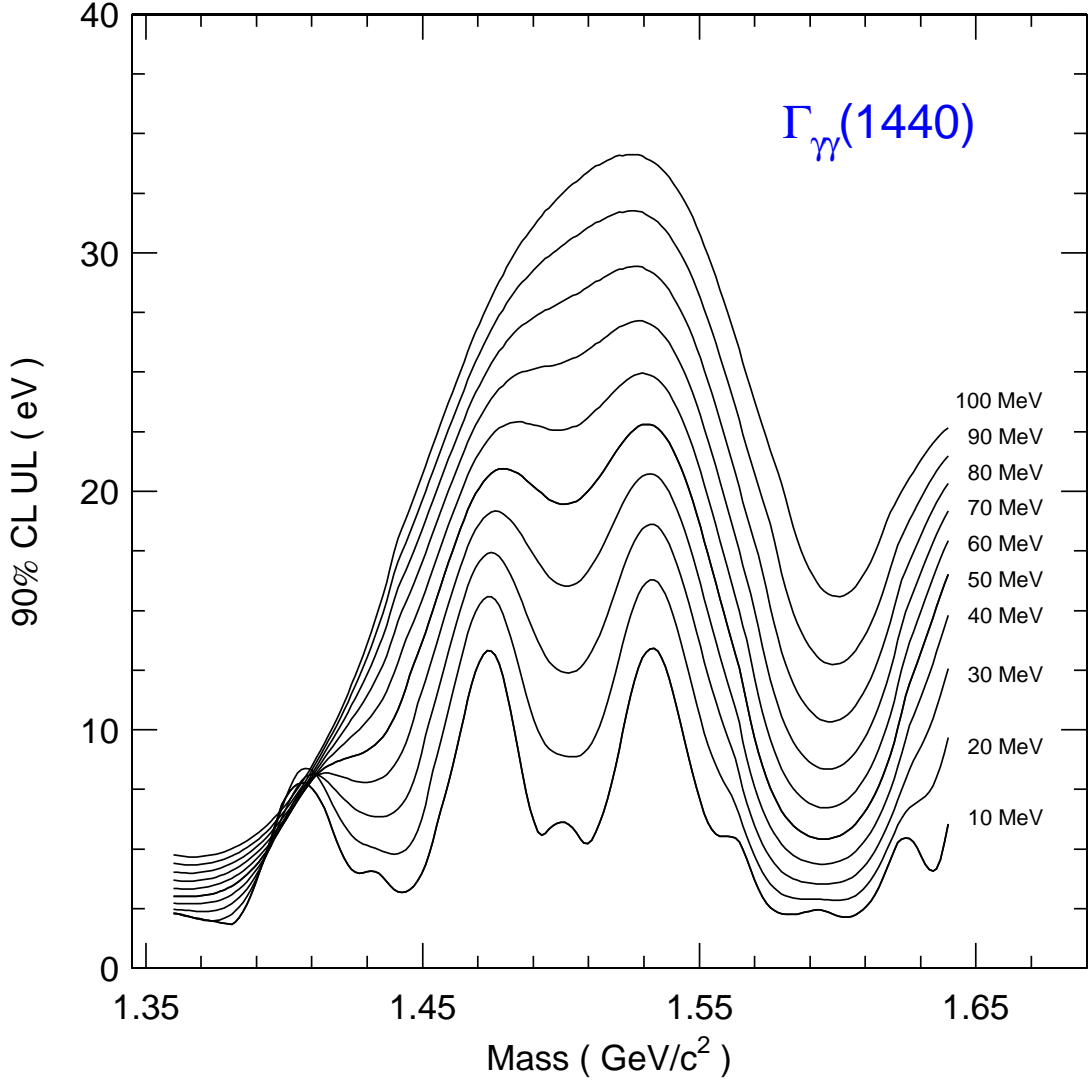}
\caption
{
90\%~CL upper limits (in eV) on 
$\Gamma_{\gamma\gamma}(1440) {\cal B}(K^0_S (\pi^+\pi^-) K^\pm \pi^\mp )$ 
versus the mass of $\eta(1440)$ 
for various hypotheses for its full width assuming 
three body phase-space decay to $K^0_S K^\pm \pi^\mp$. 
}
\label{fig_uls_1}
\end{figure}

To estimate the upper limit on the two-photon partial width 
of the $\eta(1440)$, we divide our 90\%~CL upper limit on 
the number of signal events by the product 
of the overall detection efficiency (0.76\%) 
reduced by one unit of systematic 
uncertainty (30\%), 
the integrated $e^+e^-$ luminosity 
(13.8${\rm ~fb^{-1}}$) and our 
numerical prediction for the pseudoscalar 
two-photon cross section (33~pb/keV) 
(see Eq.~\ref{equation_1}). 
The efficiency quoted above assumes the L3 mass and full width 
for the $\eta(1440)$ and its phase-space decay to $K \bar{K} \pi$. 
The high inefficiency arises primarily from the kinematics of 
two-photon collisions. 
Combining all these numbers, we obtain 
a Bayesian upper limit of 
$\Gamma_{\gamma\gamma}(1440) {\cal B}(K^0_S (\pi^+\pi^-) K^\pm \pi^\mp ) < 20.4{\rm ~eV}$ 
(at 90\%~CL), which is to be compared 
to the L3 value of $49 \pm 12 {\rm ~(stat.)}$~eV (with 11\% systematic error). 
We repeat these estimates for various hypotheses for $M(1440)$ and $\Gamma(1440)$ 
and show our results in Fig.~\ref{fig_uls_1}. 

\begin{figure}
\includegraphics[width=6.5in]{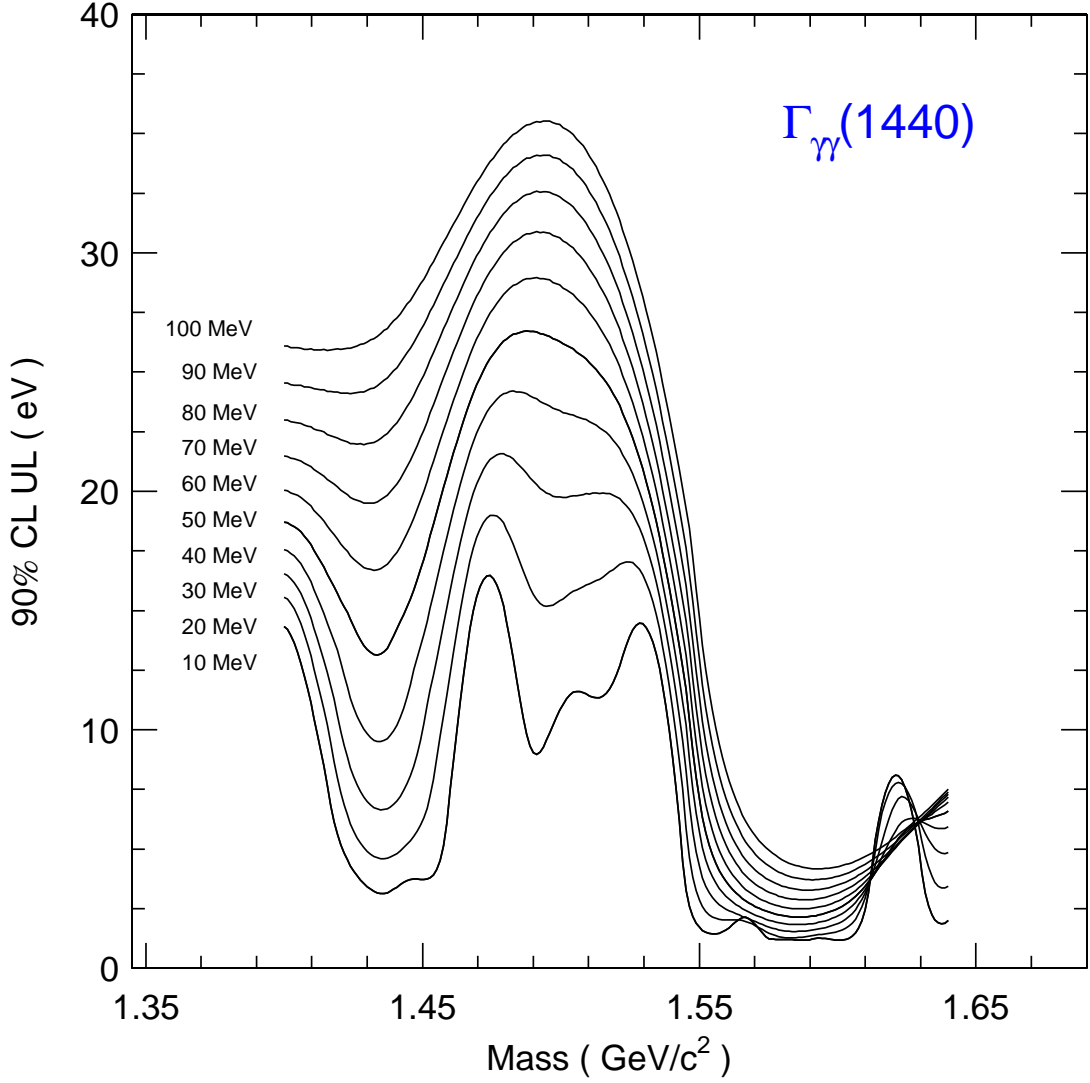}
\caption
{
90\%~CL upper limits (in eV) on 
$\Gamma_{\gamma\gamma}(1440) {\cal B}(K^0_S (\pi^+\pi^-) K^\pm \pi^\mp )$ 
versus the mass of $\eta(1440)$ 
for various hypotheses for its full width assuming 
two body decay to $\bar{K}^{*}K$ followed by the decay $K^0_S K^\pm \pi^\mp$. 
}
\label{fig_uls_2}
\end{figure}

It is possible that the $\eta(1440)$ predominantly decays into $\bar{K}^{*}K$ followed 
by the decay $K^0_S K^\pm \pi^\mp$. We study this scenario by imposing an additional 
selection criterion on the $K^0_S \pi^\mp$ or $K^\pm \pi^\mp$ invariant mass, 
requiring at least one combination to be within the $K^*$ mass window 
0.84--0.94 ${\rm GeV/c^2}$. After applying this 
requirement, we repeat the analysis described above 
and arrive at the 90\%~CL upper limits shown in Fig.~\ref{fig_uls_2}. 
To make the estimates shown in this figure, we properly account for 
the $K^*$ polarization because it affects the reconstruction efficiency. 

In addition to the lack of an $\eta(1440)$ signal, 
we do not observe any events in the $\eta(1295)$ mass region and 
use the upper limit of 2.3 signal events (at 90\%~CL) to estimate the production rate. 
This translates into an upper limit of $\Gamma_{\gamma\gamma}(1295) {\cal B}(K \bar{K} \pi) < 14$~eV 
at 90\%~CL using the mass and full width obtained by the previous experiments\cite{PDG}. 
To convert our results into upper limits on $\Gamma_{\gamma\gamma} {\cal B}(K\bar{K}\pi)$ 
we use ${\cal B}(K^0_S \to \pi^+\pi^-) = 0.686$ and 
${\cal B}(K\bar{K}\pi \to K^0_S K^\pm \pi^\mp) = 1/3$. 

We test our analysis technique by estimating the two-photon partial 
width of the $\eta_c$ that we reported recently in an independent 
analysis of the same data sample\cite{etacprime}. That analysis measures 
$\Gamma_{\gamma\gamma}(\eta_c) = 7.5 \pm 0.5 {\rm ~(stat.)} \pm 0.5 {\rm ~(sys.)}
{\rm ~keV}$, while our estimate is 
$\Gamma_{\gamma\gamma}(\eta_c) = 7.0 \pm 0.7 {\rm ~(stat.)} {\rm ~keV}$. 
Our estimates, performed independently with the CLEO~II and CLEO~II.V 
data samples are 
$7.1 \pm 1.3 {\rm ~(stat.)} {\rm ~keV}$ 
and  
$7.0 \pm 0.9 {\rm ~(stat.)} {\rm ~keV}$, 
respectively. Our larger statistical error is due to the restrictive nature of 
our unmatched neutral energy and transverse momentum criteria. 

The dominant sources of systematic error 
in the overall efficiency 
are the uncertainties in the 
four-track event selection (21\%), 
trigger efficiency (14\%), 
the unmatched neutral energy requirement (10\%), 
and the transverse momentum (10\%) requirement,
all of which are added in quadrature. 
The 21\% uncertainty in four-track selection 
consists of two 15\% uncertainties. The first 15\% error 
reflects the uncertainty in the detector simulation 
of efficiency loss for charged tracks at small 
polar angles (with respect to the beam direction). 
We evaluate this error by comparing 
the measurements of efficiency loss 
in our calibration data sample 
and in 
signal MC as a function of polar angles for the final state charged particles. 
The data and MC agreed, but we decide to use the statistical 
uncertainty on the comparison as our systematic error, which is 15\%. 
The other 15\% uncertainty is due to requiring exactly four charged tracks 
reconstructed in event candidates. To estimate the uncertainty associated with this 
selection criterion, we allow more than four charged tracks in our calibration data sample 
and in signal MC. Then we measure the efficiency of four charged tracks selection criterion. 
Again, the two samples agree on the efficiency, but we take the statistical 
uncertainty of the comparison, which is dominated by the data statistics. 
The uncertainty in the trigger efficiency is primarily due to 
imperfect knowledge of its value for events with a small 
number of charged hadrons. To estimate this uncertainty, 
we measure the trigger efficiencies in data 
using partially independent trigger condition configurations. 
These efficiencies are implemented in our detector simulation 
on a particle-by-particle basis. However, 
the deficiencies of our GEANT~3-based MC program in simulating 
hadronic interactions of charged hadrons in the calorimeter 
result in a 14\% uncertainty in the trigger efficiency for four-track events. 
The uncertainties in the unmatched neutral energy and transverse momentum 
selection criteria are determined by the statistics of our calibration data sample. 
For these two criteria, there is also a good agreement between the data 
and our two-photon MC for low-mass $K^0_S$ pairs. 

Our upper limits for $\eta(1440)$ production are inconsistent 
with the first observation of this resonance by the L3 experiment 
at a level of greater than $2~\sigma$. 
To estimate our upper limits, we have to make some assumptions about 
the invariant mass shape for background contribution.  
We do not take into account the effect of possible interference between resonant 
and continuum two-photon production of $K^0_S K^\pm \pi^\mp$. 

\section{Axial-Vector Mesons $f_1(1285)$ and $f_1(1420)$}
\label{results_axial-vectors}

\begin{figure}
\includegraphics[width=6.5in]{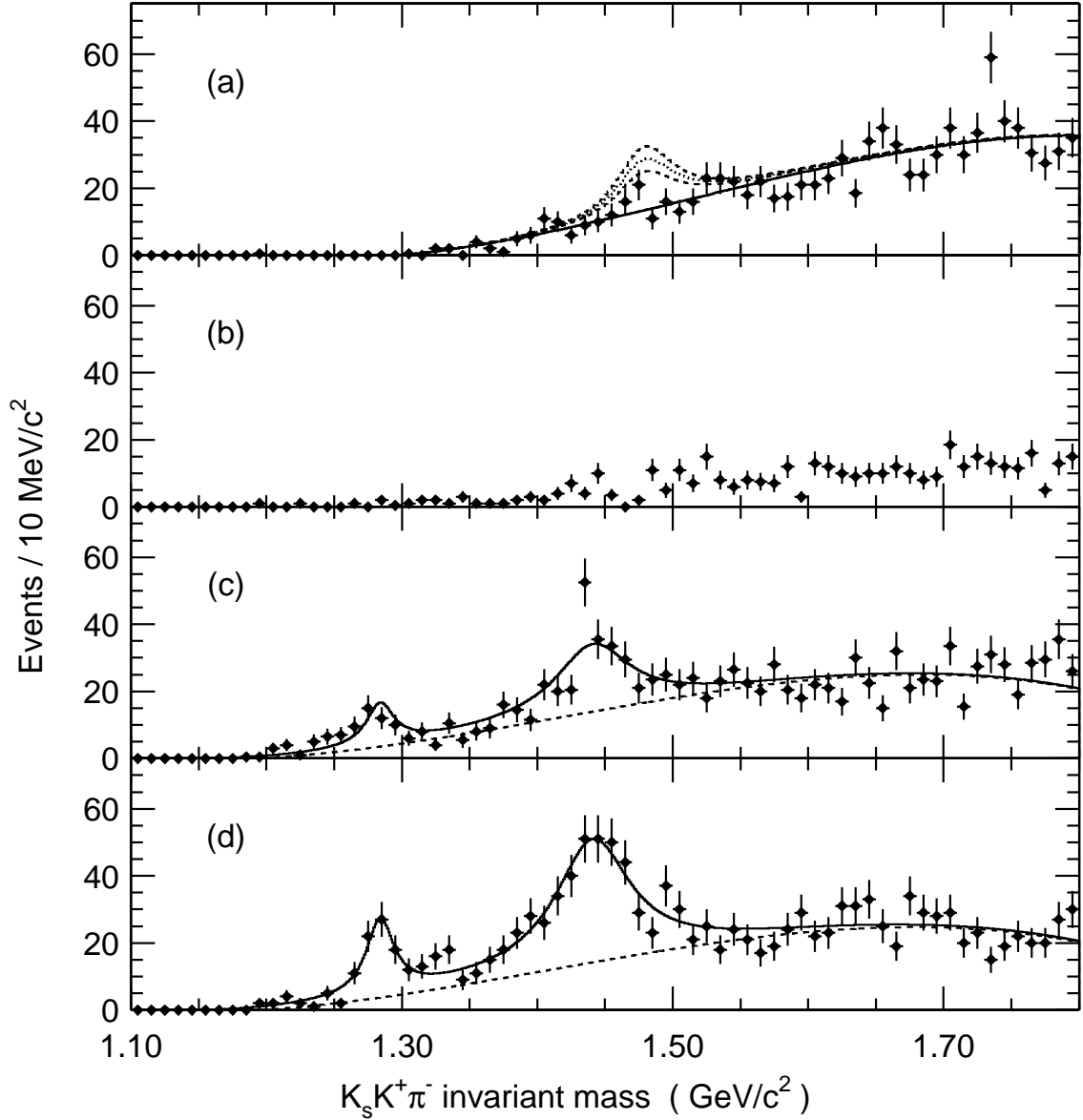}
\caption
{
The distributions of the $K^0_S K^\pm \pi^\mp$ invariant mass for data events detected with 
(a) $p_\perp \le 100 {\rm ~MeV/c}$, 
(b) $100 {\rm ~MeV/c} \le p_\perp \le 200 {\rm ~MeV/c}$ and 
(c) $200 {\rm ~MeV/c} \le p_\perp \le 600 {\rm ~MeV/c}$ 
in the untagged mode, and 
(d) for all $p_\perp$ in the tagged mode. 
The dashed curves in (a) show the strength of the expected $\eta(1440)$ signal 
according to the L3 results\protect\cite{L3}. 
The solid curves in (a), (c) and (d) are the results of binned maximum likelihood fits 
for resonances with a polynomial approximation to the non-interfering combinatorial background. 
For (a), only the strength of the $\eta(1440)$ resonance is fit.  
For (c) and (d), the strength of $f_1(1285)$ and $f_1(1420)$ are fit. 
For the fit shown in (a), we use the mass and full width according to L3 estimates\protect\cite{L3}. 
For the fit shown in (c), we use the masses and full widths obtained from our fit 
to the tagged invariant mass distribution shown in (d). 
}
\label{fig_data_mass_kskp_7}
\end{figure}

After observing no pseudoscalar signals at low $p_\perp$, we inspect 
the distributions of $K^0_S K^\pm \pi^\mp$ invariant mass in untagged events 
at larger $p_\perp$. These distributions are shown in Figs.~\ref{fig_data_mass_kskp_7}(b)~and~(c). 
At intermediate $p_\perp$, shown in Fig.~\ref{fig_data_mass_kskp_7}(b), 
we see no indication of $\eta(1295)$ and $\eta(1440)$ 
which is consistent with our non-observation of these hadrons 
at lower transverse momentum. 
However, Fig.~\ref{fig_data_mass_kskp_7}(c) indicates two enhancements for 
$K^0_S K^\pm \pi^\mp$ events with $200 {\rm ~MeV/c~} < p_\perp < 600{\rm ~MeV/c}$. 
Fitting this untagged high-$p_\perp$ mass distribution to the $\eta(1295)$ 
(using the nominal mass and width\cite{PDG}),
and to the $\eta(1440)$ (using L3's mass and width) 
on top of a third order polynomial for the 
combinatorial background\footnote{
For the fit shown in Fig.~\ref{fig_data_mass_kskp_7}(c), 
we use the masses and the widths we obtain from 
our fit to the tagged mass distribution 
shown in Fig.~\ref{fig_data_mass_kskp_7}(d).}, 
we find $214 \pm 33$ events in the higher mass enhancement and 
$48 \pm 13$ events in the lower mass enhancement. 
We investigate the hypothesis that our observations are due to 
statistical fluctuations of untagged $\eta(1440)$ signal events, 
which should populate the low-$p_\perp$ region, into 
the high-$p_\perp$ region 
between 200~MeV/c and 600~MeV/c. We perform 10,000 toy MC experiments 
to estimate the probability for 200 pseudoscalar two-photon events to fluctuate 
to this $p_\perp$ region in the presence of the background observed in data. 
In each toy MC experiment, we use the shape of the invariant mass 
distribution for background and the number of background events estimated from data 
in the high $p_\perp$ region. In every run we also generate 200 signal toy MC events 
with the predicted $p_\perp$ distribution. 
These signal toy MC events are then merged with the toy background sample. 
Then, we perform a binned ML fit to the mass distribution 
for the high $p_\perp$ region and estimate the central value 
for the number of signal $\eta(1440)$ events. 
All our toy MC experiments yield less than 120 $\eta(1440)$ events, 
with a peak in the distribution of signal yield at $55 \pm 16$ events. 
Therefore, we estimate the signal fluctuation probability to be less than $10^{-4}$. 
This is not surprising because 
transverse momentum distributions 
for two-photon MC (shown in Fig.~\ref{fig_mc_ptran_1})
and 
calibration two-photon data (shown in Fig.~\ref{fig_data_ptran_ksks_3}) 
peak sharply towards small values. 

We use previous measurements from TPC/2$\gamma$\cite{tpc2g_f1} 
and our Monte Carlo generator for two-photon axial-vector production 
to estimate the expected rate for the $f_1(1420)$ signal 
in the high-$p_\perp$ region (between 200~MeV/c and 600~MeV/c) 
to be $200 \pm 50$ events, 
which is consistent with the observed $214 \pm 33$ events. 
Also, the enhancement at lower invariant mass in data, 
shown in Fig.~\ref{fig_data_mass_kskp_7}(c), 
is consistent with the $f_1(1285)$ hypothesis. 
These two mesons, $f_1(1285)$ and $f_1(1420)$, 
are the only axial-vectors 
consistent with quark model predictions. 
However, because our untagged events contain at least one missing 
particle, it is impossible to rule out backgrounds from some other processes, 
such as hadronic decays of $\tau$ leptons or threshold enhancements in 
two-photon production of $K^*\bar{K}^* n(\pi)$.\footnote{These 
backgrounds also complicate the analysis 
of two-photon resonances in the low-$p_\perp$ region, 
but to a lesser degree than 
in untagged two-photon events of higher $p_\perp$.} 
Therefore, we cannot claim the observation of two-photon 
production of axial-vector mesons solely on the basis of the 
untagged invariant mass distribution shown in 
Fig.~\ref{fig_data_mass_kskp_7}(c).

\begin{figure}
\includegraphics[width=6.5in]{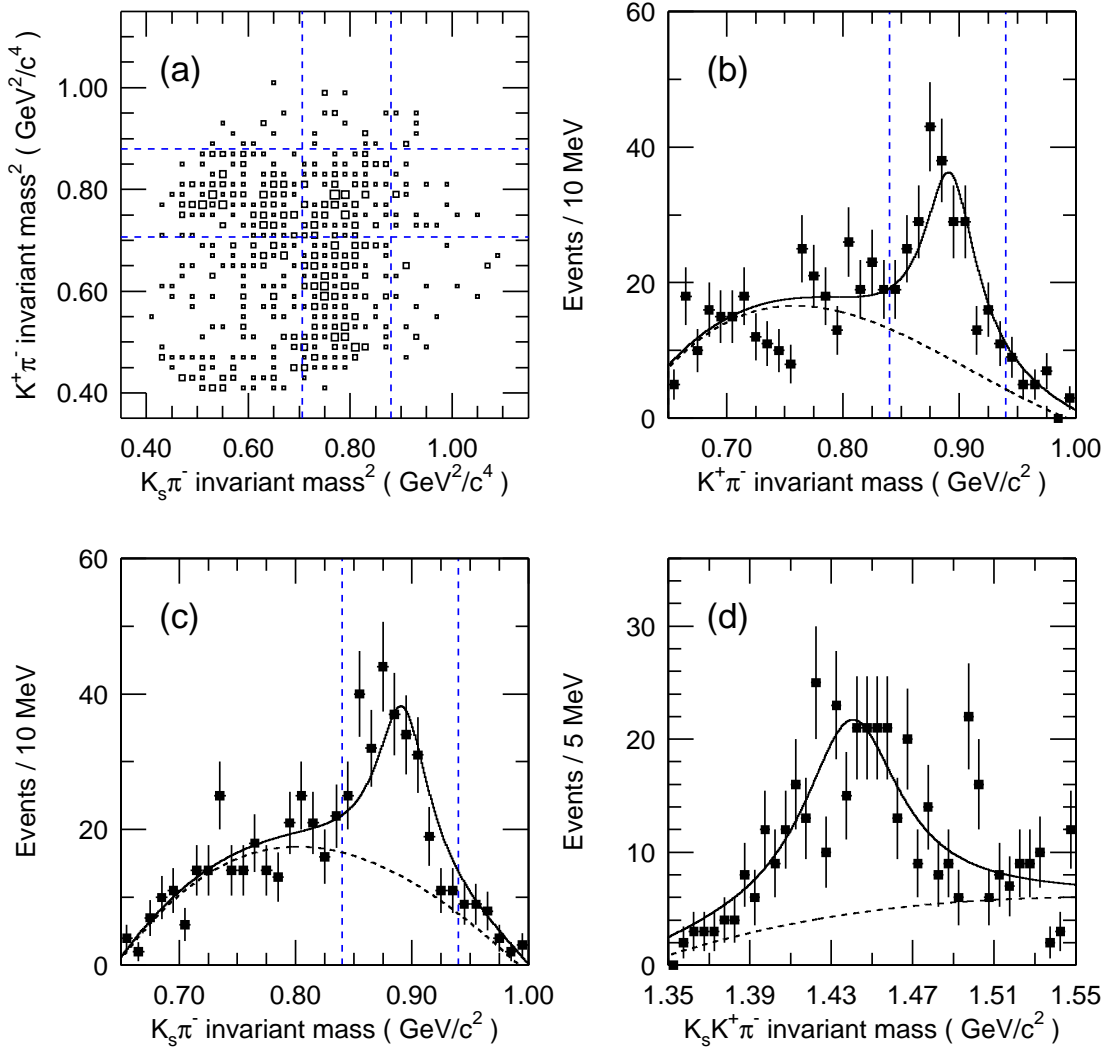}
\caption
{
Dalitz plot for $f_1(1420)$ candidates detected 
in tagged data events in the $K^0_S K^\pm \pi^\mp$ invariant mass region 
between $1.35{\rm ~GeV/c^2}$ and $1.55{\rm ~GeV/c^2}$ (a). 
The projections of Dalitz plot on the $K^\pm \pi^\mp$ (b) 
and $K^0_S \pi^\mp$ (c) invariant masses. 
The $K^0_S K^\pm \pi^\mp$ invariant mass distribution (d) 
for data events with at least one $K \pi$ combination in the 
$K^*$ mass region indicated by the vertical and horizontal 
bands in (a), (b) and (c). 
Solid and dashed curves show the results of the fits. 
}
\label{fig_data_tagged_dalitz_6}
\end{figure}

To unambiguously establish the presence of axial-vector mesons in our data, 
we use the distribution of the invariant mass for tagged two-photon events 
shown in Fig.~\ref{fig_data_mass_kskp_7}(d). 
The statistics shown in this figure allow 
us to perform ML fits for the masses and the widths of both enhancements. 
Our results are $M = 1284 \pm 3 {\rm ~MeV/c^2}$ and $\Gamma = 25 \pm 6$~MeV for 
the lower mass enhancement, and 
$M = 1441 \pm 3 {\rm ~MeV/c^2}$ and $\Gamma = 67 \pm 9$~MeV for the higher mass enhancement. 
This is consistent with the results of previous 
experiments\cite{PDG} for the $f_1(1285)$ and $f_1(1420)$. 
From our fits, we conclude that both invariant mass enhancements in 
Figs.~\ref{fig_data_mass_kskp_7}(c)~and~(d) 
are due to axial-vector two-photon production. 
The $f_1(1420)$ assignment is supported by the strong signal 
for $K^{*}$ decays observed in the Dalitz plot for the $f_1(1420)$ 
signal region in tagged data events, 
shown in Fig.~\ref{fig_data_tagged_dalitz_6}(a). 
According to previous experiments\cite{PDG}, the $f_1(1420)$ 
indeed decays dominantly into the $\bar{K}^{*}K$ channel. 
The projections of the Dalitz plot are shown in 
Figs.~\ref{fig_data_tagged_dalitz_6}(b)~and~(c), 
the $K^0_S K^\pm \pi^\mp$ invariant mass distribution 
for the $f_1(1420)$ signal region is shown 
in Fig.~\ref{fig_data_tagged_dalitz_6}(d). To be included 
in the latter distribution, at least one $K \pi$ combination 
is required to be in the $K^*$ signal region indicated 
by the vertical and horizontal bands 
in Figs.~\ref{fig_data_tagged_dalitz_6}(a),(b)~and~(c). 
In our fits, we approximate the resonances by 
two-body relativistic Breit-Wigners 
ignoring (small) resolution effects. 

Our previous two-photon publication\cite{paper_ff} demonstrated that, 
with CLEO, 
we can measure the invariant masses of light mesons with better than 
a few MeV accuracy. It is interesting to note that the enhancement 
we associate with the $f_1(1420)$ has its mass shifted relative to the nominal value 
of $1426.3 \pm 0.9 {\rm ~MeV/c^2}$ by more than four standard deviations 
while the width is consistent with $54.9 \pm 2.6$~MeV\cite{PDG}. 
This discrepancy could be due to 
the interference with non-resonant two-photon continuum 
production that we do not take into account in our analysis. 
There is less non-resonant background in the $f_1(1285)$ 
mass region, and our estimates are indeed in a better agreement with the 
nominal values\cite{PDG} of $M = 1281.8 \pm 0.6 {\rm ~MeV/c^2}$ and $\Gamma = 24.1 \pm 1.1$~MeV. 
Our $K^0_S K^\pm \pi^\mp$ invariant mass distributions show no indication of the $f_1(1510)$. 

\section{Conclusions}
\label{conclusions}

With a data sample that exceeds the L3 statistics by a factor 
of five, we cannot confirm their first observation 
of the $\eta(1440)$ in two-photon collisions. 
We report the upper limits 
$\Gamma_{\gamma\gamma}(1440) {\cal B}(K\bar{K}\pi) < 89~{\rm eV}$ 
(with the mass $1481 ~{\rm MeV/c^2}$ and full width $48 {\rm ~MeV}$  
according to L3 estimates\cite{L3}) 
and 
$\Gamma_{\gamma\gamma}(1295) {\cal B}(K\bar{K}\pi) < 14~{\rm eV}$ 
(assuming nominal mass and full width\cite{PDG}) 
at 90\%~CL. The former number should be compared 
with the result reported by the L3 experiment 
$\Gamma_{\gamma\gamma}(1440) {\cal B}(K\bar{K}\pi) = 212 \pm 50 {\rm ~(stat.)} \pm 23 {\rm ~(sys.)}$. 
Our results for the $\eta(1440)$ are more than 
two standard deviations inconsistent 
with L3's measurements.
Our upper limits on the two-photon partial widths of 
the $\eta(1295)$ and $\eta(1440)$ are consistent with the glueball and 
the radial excitation hypotheses. 

We do not observe any pseudoscalar mesons with masses 
below $1.7~{\rm GeV/c^2}$ of sizable $\Gamma_{\gamma\gamma}$ 
decaying dominantly to $K\bar{K}\pi$. 
We observe only two ground-state axial-vector mesons in this mass region,  
consistent with quark model expectations. 
If the third axial-vector meson, the $f_1(1510)$, does exist, 
it may be a glueball or an exotic particle. 
In principle, it is possible that mesons and glueballs mix in nature. 
However, our observation of no extranumeral pseudoscalars and 
of exactly two axial-vector mesons decaying to $K\bar{K}\pi$ 
makes such an interesting but somewhat artificial scenario unnecessary 
to introduce for light hadrons decaying in this channel. 

We expect to obtain definitive information 
on the existence and the origin of light pseudoscalar exotic 
hadrons decaying to $K\bar{K}\pi$ with our currently 
running CLEO-c experiment\cite{cleoc}, where 
the analysis of radiative hadronic decays 
of $\sim 1$ billion $J/\psi$'s might result 
in the firm discovery of glueballs and light exotics. 
We gratefully acknowledge the effort of the CESR staff 
in providing us with excellent luminosity and running conditions.
This work was supported by the National Science Foundation
and the U.S. Department of Energy.

\end{document}